# Thermal response of double-layered metal films after ultrashort pulsed laser irradiation: the role of nonthermal electron dynamics

George D. Tsibidis[♣]

*Institute of Electronic Structure and Laser (IESL), Foundation for Research and Technology (FORTH), N. Plastira 100, Vassilika Vouton, 70013, Heraklion, Crete, Greece*

The thermal response of a Cu-Ti double-layered film is investigated after laser irradiation with ultrashort pulses (pulse duration $\tau_p$=50fs, 800nm laser wavelength) in submelting conditions by including the influence of nonthermal electrons. A revised two-temperature model is employed to account for the contribution of nonthermal electron distribution while the variation of the optical properties of the material during the laser beam irradiation is also incorporated into the model. Theoretical results can provide significant insight into the physical mechanism that characterize electron dynamics and can facilitate production of controllable ultra-high strength Cu-Ti alloys with promising applications.

---

[♣] E-mail: tsibidis@iesl.forth.gr

Research in double-layered metal thin films (DLMTF) has received considerable attention over the past decade due to their important technological applications, in particular in photo/micro-electronic devices [1] and micro-electro-mechanical switches (MEMS) [2,3]. One highly promising material that has been suggested as an ultra-high strength and high conductivity material for applications such as production of conductive springs, interconnections and connectors in photovoltaics is a Cu-Ti alloy that aims to displace existing Cu–Be alloys. This proposal has been dictated by serious health hazards associated with the Be-based metallurgy in production [4].

To investigate the ultrafast heating characteristics, DLMTF assemblies are irradiated with ultrashort pulsed lasers [3,5]. The exploitation of heat capacity and electron-phonon coupling coefficient differences between the constituent layers aim to induce reduction of the surface temperature [5], increase of the damage threshold (i.e. melting onset) [6] and eventually enhancement of the mechanical strength of the upper layer [7]. A common approach to simulate energy transfer and relaxation is the traditional two temperature model (TTM) [8]. Nevertheless, due to the fact that nonthermal electron interactions are not incorporated in the theoretical framework, the model is expected to be inadequate to describe accurately the physical mechanism in the first femtoseconds after the excitation. Various approaches were proposed based on the Boltzmann transport equation aimed at overcoming that limitation and describing efficiently the nonthermal electron dynamicsb [9-12] with promising agreement with experimental observations. Alternative approaches based on the introduction of a three temperature model to incorporate nonthermal electron baths yielded excellent agreement between theoretical and experimental results [12,13], however, they required the inclusion of fitting constants while they still have not been employed to explore electron dynamics in multi-layered films. Furthermore, one important issue that is commonly ignored is that reflection and absorption coefficients are assumed to remain constant during lasing which yields inaccurate estimation of optical properties and the lattice and electron temperatures [14].

In this letter, we propose an extended TTM model (ETTM), which, in comparison to previous approaches that include the nonthermal electron contribution [15,16], comprises transient changes of optical characteristics during the irradiation with femtosecond pulse lasers to describe ultrafast dynamics and thermal response in Cu-Ti films. To compute the temperature dependence of the absorption and reflection coefficients, an extension of the simple Drude model [14,17,18] is considered in which Lorentzian terms are included in the dielectric constant function with parameters that enhance agreement with experimental data for wavelengths ~800nm [19]. A thorough investigation of the fundamental mechanisms will allow the determination of the influence of the titanium layer that has substantially larger electron-lattice coupling. Moreover, to take into account ballistic motion of the excited electrons, absorption penetration depth is replaced with an effective depth that includes ballistic transport $\Lambda$ ( $\Lambda$=70nm for Cu [20,21]). To simplify calculations and avoid complex effects at the interface of the two metals, Cu film thickness is assumed to be large enough to allow laser beam energy dissipation without penetrating the Ti layer.

A schematic diagram is illustrated in Fig. 1, in which a DLMTF made up of Cu and Ti with thicknesses $d_1$ and $d_2$, respectively, is irradiated with an ultrashort pulsed (polarised along x-axis and on the plane of incidence) Gaussian laser beam (pulse duration $t_p$=50fs, laser beam fluence $E_p$=0.06J/cm$^2$, 800nm laser wavelength, irradiation spot radius $R_0$=10μm) that propagates along the *z*-axis. Laser beam characteristics have been chosen so that lattice temperature in both materials will not exceed the melting point ($T_m(Cu)$=1357K, $T_m(Ti)$=1941K).

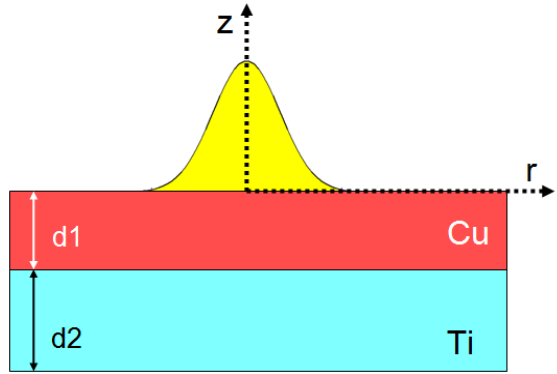

FIG. 1. (Color Online) Schematic of a Cu-Ti DLMTF irradiated with a 50fs Gaussian laser beam ($E_p$=0.06J/cm$^2$, 800nm laser wavelength, $R_0$=10μm).

To describe the influence of the ultrafast electron dynamics in the relaxation procedure after irradiation of the DLMTF in submelting conditions, heat transfer of the nonthermal electron distribution to the thermal electrons and lattice is included [15], along with the thermal electron-lattice interaction [6,22,23]. Hence, the following set of equations is employed to investigate the spatio-temporal distribution of electron ($T_e$) and lattice (T$_L$) temperatures of the assembly

$$C_e^{(1)} \frac{\partial T_e^{(1)}}{\partial t} = \vec{\nabla} \cdot \left( k_e^{(1)} \vec{\nabla} T_e^{(1)} \right) - G_e^{(1)} \left( T_e^{(1)} - T_L^{(1)} \right) + \frac{\partial U_{ee}}{\partial t}$$

$$C_L^{(1)} \frac{\partial T_L^{(1)}}{\partial t} = G_L^{(1)} \left( T_e^{(1)} - T_L^{(1)} \right) + \frac{\partial U_{eL}}{\partial t}$$

$$C_e^{(2)} \frac{\partial T_e^{(2)}}{\partial t} = \vec{\nabla} \cdot \left( k_e^{(2)} \vec{\nabla} T_e^{(2)} \right) - G_e^{(2)} \left( T_e^{(2)} - T_L^{(2)} \right) \quad (1)$$

$$C_L^{(2)} \frac{\partial T_L^{(2)}}{\partial t} = G_L^{(2)} \left( T_e^{(2)} - T_L^{(2)} \right)$$

where the subscripts $e$ and $L$ are associated with electrons and lattice, respectively, $k_e$ is the thermal conductivity of the electrons, $C_e$ and $C_L$ are the heat capacity of electrons and lattice, respectively, $G_L$ is the electron-phonon coupling constant while superscripts $i$ correspond to the Cu ($i$=1) and Ti layers ($i$=2), respectively. The energy densities per unit time transferred from the nonthermal electrons to thermal electrons ($\partial U_{ee}/\partial t$) and lattice ($\partial U_{eL}/\partial t$) require modification with respect to the initial model [15] to account for the dynamic character of the reflectivity and the absorption coefficients during irradiation that alters the absorption and the laser heat density distribution, respectively.

To take into account temperature dependence of the optical characteristics, the dielectric constant of Cu is modelled by means of an extended Lorentz-Drude model with three Lorentzian terms [14,17] to enhance agreement with experimental data [19]

$$\varepsilon(\omega) = \varepsilon_\infty - \frac{(\omega_D)^2}{\omega^2 + i\gamma\omega} + \sum_{p=1}^{3} B_p \Omega_p \left( \frac{e^{i\phi_p}}{\Omega_p - \omega - i\Gamma_p} + \frac{e^{-i\phi_p}}{\Omega_p + \omega + i\Gamma_p} \right) \quad (2)$$

where $\varepsilon_\infty$ is dielectric constant at infinite frequency, $\omega_D$ is the plasma frequency, $\omega$ is the laser frequency (=$2.3562 \times 10^{15}$ rad/s), $\gamma$ equals the reciprocal of electron relaxation time, $B_p$ is a weighting factor, while $\varphi_p$, $\Gamma_p$, and $\Omega_p$ are the phase, gap, and broadening, respectively, where values of the aforementioned parameters are provided in the Supplementary Material (SM) [24]. Hence, the spatio-temporal values of the reflection $R(r,z=0,t)$ on the Cu surface and absorption coefficient $\alpha(r,z,t)$ can be determined by the Fresnel functions [24,25] which after the inclusion of the ballistic electron transport correction [20] lead to the following compact form for the expressions giving the source terms in Eq.1

$$\frac{\partial}{\partial t}\begin{Bmatrix}U_{ee}\\U_{eL}\end{Bmatrix} = \frac{2\sqrt{\ln 2}E_p}{\sqrt{\pi}(h\nu)^2 t_p}\int_0^t \left[(1-R(r,z=0,t'))\frac{1}{1-\exp\left(-\frac{d_1}{\alpha^{-1}+\Lambda}\right)}\frac{1}{\alpha^{-1}+\Lambda}\right.$$

$$\left.\times\exp\left(-4\sqrt{\ln 2}\left(\frac{t'-t_0}{t_p}\right)^2\right)\exp\left(-2\left(\frac{r}{R_0}\right)^2\right)\exp\left(-\int_0^z \frac{1}{\alpha^{-1}+\Lambda}dz'\right)\begin{Bmatrix}H_{ee}(t-t')\\H_{eL}(t-t')\end{Bmatrix}\right]dt' \quad (3)$$

where $h\nu$=1.55eV (not sufficient to allow interband transitions in Cu) and $t_0=3t_p$ is chosen so that lasing starts in practice at t=0 while the analytical expressions for $H_{ee}$ and $H_{eL}$ are provided in SM [24]. Eq.1,3 indicate that the laser irradiation, production of nonthermal electrons and interaction with electron and lattice baths, thickness of the upper layer, thermo-physical properties of the substrate and heat transfer between the two materials are all correlated; therefore the thermal response of the surface of the material will be related both to the laser-matter interaction mechanism and the influence of the substrate layer. It has also to be emphasised that throughout the presented work the thickness of the upper layer is selected to be large enough to assume negligible penetration of the laser beam into the substrate layer and therefore the source term for the second layer approximately vanishes. To propose a more complete and rigorous framework that can be employed for thinner Cu layers and include laser irradiation that penetrates the second layer, a source term would be required to be presented to describe the effect of light propagation in the substrate. Certainly, a simplistic argument is to present spatial distribution of the light intensity inside the DLMTF and use the analytic expressions of wave propagation (i.e. interference of the transmitted component and reflected wave on the interface determine the net energy deposition in the first film while a transmitted wave propagates inside the substrate) [26]. Experimental observations indicate a satisfactory agreement with theoretical results in which the ballistic electron contribution has been technically included to describe a new cooling mechanism and provide a correction to the effective penetration depth [20]. However, no prior experimental evidence exists that confirms the behaviour of the net movement of the ballistic electrons as a result of the interference of the incident and reflected (on the interface) waves and thereby a theoretical assumption for a potential behaviour may be unrealistic.

The numerical solution of Eqs.1-3 was performed using a finite difference method scheme where it is assumed that there exists perfect thermal contact between the two metals, and electron temperature and heat flux is continuous on the interface, thus interface resistance is neglected. The thickness of the second layer is assumed to be large enough that laser beam will be attenuated before it reaches the back side of the film; this allows to assume von Neumann boundary conditions and ignore heat losses at the front and back surfaces of the assembly. To simplify the two-dimensional solving technique, a coordinate system in cylindrical coordinates is employed to solve Eqs.1-3 assuming an axial symmetry while the mesh discretisation parameters are $\Delta t$=0.1fs, $\Delta r$=20nm, $\Delta z$=1nm.

Fig.2a illustrates the evolution of the absorption coefficient and the reflectivity where the transient electron and lattice temperatures values that are used to evaluate the dielectric constant of Cu are computed employing TTM. Calculations are shown at $(r,z)=(0,0)$ while a similar behaviour is exhibited in other parts of the material (results not shown). It is evident that both the reflectivity and absorption coefficient that influence the amount of energy deposition and spatial distribution, respectively, inside the DLMTF vary substantially within the pulse duration. Hence, the assumption of constant reflectivity and absorption coefficient which is normally assumed to describe transient dynamics [20,22] during the pulse duration would lead to incorrect results. Furthermore, according to Fig.2a, the dielectric constant variation leads to an increase of the optical penetration depth and an increase of the amount of absorbed energy.

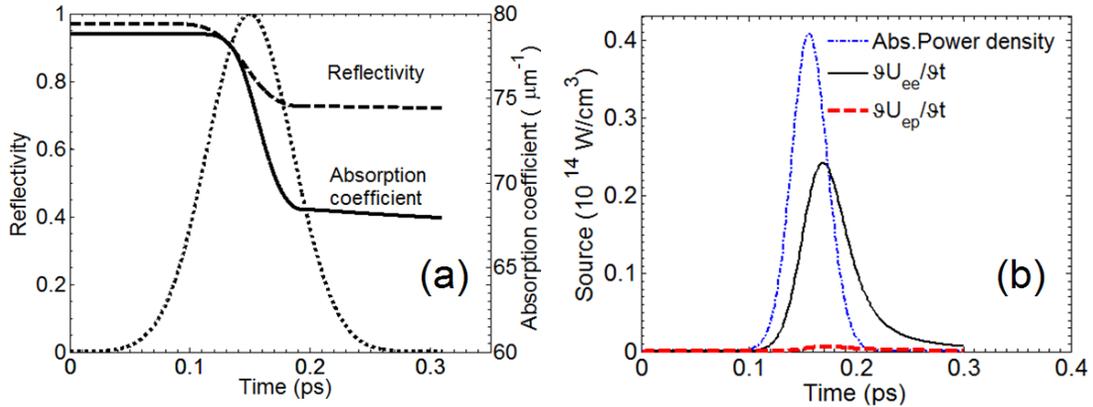

FIG. 2. (Color Online) (a) Reflectivity and absorption coefficient variations and (b) Time evolution of the heat sources for: (i) the absorbed power density (*dashed-dotted* line), (ii) $\partial U_{ee}/\partial t$ (*solid* line) and (iii) $\partial U_{el}/\partial t$ (*dotted* line) at $(r,z)=(0,0)$ for Cu irradiated with a 50fs Gaussian laser beam ($E_p$=0.06J/cm$^2$, 800nm laser wavelength, $R_0$=10μm, $d_1$=100nm).

To determine the role of the nonthermal electrons in the heat exchange and relaxation process, it is important to provide a computed estimate of the energy density per unit time stored in the nonthermal electronic distribution. Fig.2b illustrates the power density which in the traditional TTM is stored in the thermal electron system while the net result of the contribution of $\partial U_{ee}/\partial t$ and $\partial U_{eL}/\partial t$ indicates the large amount of overestimation of the internal energy of the thermal electrons. It is evident from the temporal profile of the lattice heat source that electron-electron scattering dominates in the initial stages due to the small electron-phonon coupling.

A comparison of the surface electron and lattice temperatures at $r=0$ as a function of time (for $t_p$=50fs, $d_1$=100nm, $d_2$=300nm) simulated with ETTM and TTM is presented in Fig. 3a. It is evident from viewing and comparing the two models that the electron temperature peak is reached with a short delay with respect to the value attained if the TTM model is employed (this behaviour is also illustrated in Fig.3b). This consequence is ascribed to the erroneous assumption of an instantaneous creation of thermal electrons and heating process of the electron and lattice baths predicted from the TTM model; in practice, there is a finite time required for the creation of the nonthermal electrons which leads to a delayed heating of both the thermal electrons and lattice. The employment of the revised model indicates that the incorporation of the contribution of the nonthermal electrons and decrease of the values of the optical properties within the pulse duration lowers substantially the electron temperature and the change is more profound if results from ETTM are compared to calculations from the dynamic TTM (Fig.3a). The decrease of the electron temperature is a physical outcome of the electron energy loss due mainly to the scattering of the nonthermal electrons from the electronic bath. By contrast, the small $\partial U_{eL}/\partial t$ and secondly the large lattice heat capacity are not sufficient to produce an equally remarkable variation in the increase of the lattice temperature as also noticed in previous works [15]. Nevertheless, the inclusion of the change of the optical characteristics during the pulse leads to remarkable changes (Fig. 3a). According to Fig.3a, the traditional TTM with constant optical properties throughout the pulse duration would lead to a substantial underestimation of the surface temperature as the energy deposition change via the reflectivity variation influences more the heat exchange mechanism than the distribution of energy via the increase of the optical penetration depth. Compared to the revised TTM model (with the inclusion of the optical properties variation), ETTM results into a small rise of the lattice temperature due to small value of $\partial U_{eL}/\partial t$ but substantial change in the electron dynamic*s*. From a material science point of view, the comparison of the models suggests that the theoretically estimated value of the fluence that induces damage threshold has to be increased. This prospect has a very significant impact on material properties and industrial applicability in terms of capability to modulate laser parameters as it will provide a more accurate and precise range of fluences to avoid surface damage. It is also important to emphasise that the calculated temperature is influenced both from the laser-matter interaction mechanism and heat exchange mechanism between the two layers (Fig.3S(b) in SM [24]) which also illustrates the impact of the substrate. Furthermore, it is evident that the revised model reveals there is a very large difference in the electron dynamics (with an immediate effect to the optical properties of the material) compared to predictions from TTM.

To evaluate the pulse duration limit below which the TTM is inadequate to offer a precise mechanism description as the two models yield results with an increased discrepancy, a comparison was performed for various values of pulse duration in the range $t_p$=0.050-5ps. Fig.3b illustrates the maximum value of the maximum lattice temperature (i.e. at $r$=0, $z$=0) attained by using ETTM and TTM (with and without variation of the optical characteristics) and the resulting curves indicate the nonthermal electron role is insignificant above ~2ps which suggests that for longer pulse duration the traditional TTM yields a correct picture of the underlying physical process. Furthermore, with increasing pulse duration, the dynamical character of the optical properties disappears and therefore a static treatment is sufficient for pulses longer than ~2.1ps.

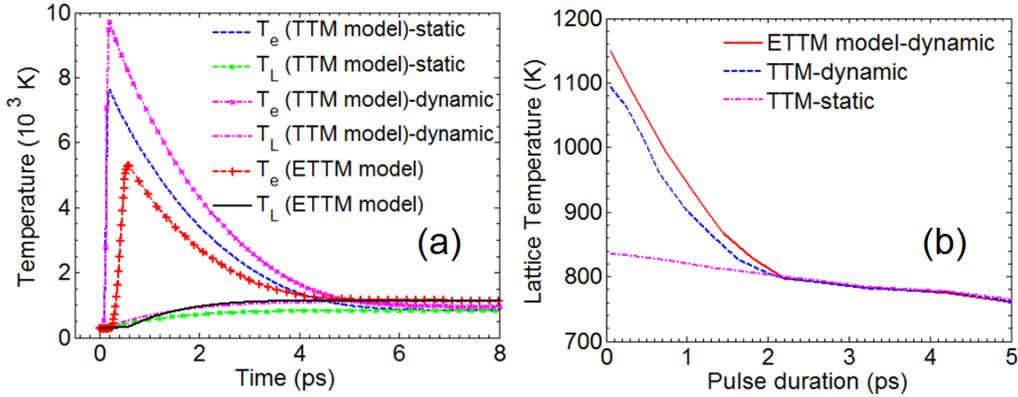

FIG. 3. (Color Online) (a) Lattice and electron temperature evolution at $(r,z)$=(0,0) calculation using ETTM and TTM models (with and without consideration of variation of the optical properties) for $t_p$=50fs, (b) Maximum Lattice temperature for various pulse duration values at $(r,z)$=(0,0). ($E_p$=0.06J/cm$^2$, 800nm laser wavelength, $R_0$=10μm, $d_1$=100nm).

Further analysis was carried out to investigate the spatial variation of the lattice temperature inside the DLMTF assembly and the particular influence of the substrate material on the thermal response of the upper layer. Fig. 4a illustrates the spatial distribution of the maximum lattice temperature at $t$=6ps which demonstrates the sharp change in the region near the interface as a result of the strong electron-phonon coupling of the substrate layer. The lattice temperature spatial variation is sketched in Fig.4b (across $r$=2.5μm for various values of Cu layer thickness ($d_1$=100, 140, 160, 200nm) at $t$=6ps. The decrease of the maximum value of the lattice temperature with increasing $d_1$ emphasises the significant influence of the heat capacity of the substrate material (i.e. $C_L(Ti) < C_L(Cu)$) in the electron-phonon relaxation process. It is important to underline the opposite thermal behaviour in DLMTFs characterised by a substrate material that has a different combination of thermophysical parameters (i.e. both larger electron-phonon coupling and phonon thermal capacity [3,23]). Furthermore, the results indicate that the influence of the electron heat conductivity ($k_e$(Cu)>> $k_e$(Ti) [27]) on the form of the lattice temperature spatial variation is not of primary importance; by contrast, the heat localization which is expressed through the electron-phonon coupling plays a predominant role as also been observed in other studies [6,28,29]. Temperature modulation by controlling the upper layer thickness and the theoretically computed change of thermal response due to the presence of the substrate could be used not only to optimize damage threshold of the DLMTF but also to determine the combination of the thickness values of the two materials that can lead to production of high-strength Cu-Ti DLMTF (due to the high tensile strength of Ti) with application-based optical and morphological properties. Furthermore, according to Fig.4b, the discrepancy of the electron-phonon coupling constants between the two materials induce the sharp lattice temperature 'jump' on the interface as the excited electrons on the first upper layer couples less effectively with lattice.

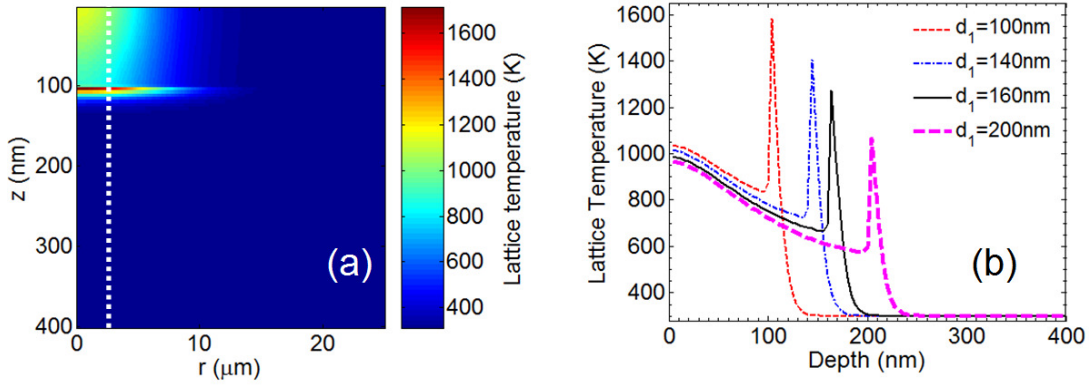

FIG. 4. (Color Online) (a) Lattice temperature field at $t$=6ps, (b) Lattice temperature spatial variation for $d_1$=100, 140, 160, 200nm across $r$=2.5μm (*white* line in (a)) at $t$=6ps. ($E_p$=0.06J/cm$^2$, $t_p$=50fs, 800nm laser wavelength, $R_0$=10μm).

In summary, a revised two temperature model has been introduced to investigate the ultrafast processes and thermal response after irradiation of DLMTF with ultrashort laser pulses in which the contribution of nonthermal electrons, material optical properties variation and the resulting impact on the thermophysical response of the assembly were explored. Simulation results demonstrate that the traditional TTM is expected to result into an underestimation of the theoretically computed damage threshold and, therefore, the proposed model could provide both the limit for validity of the conventional approach and a revised experimental damage threshold value. In terms of usability, the model aims to allow firstly to enhance our understanding of the fundamental physical mechanisms during the early stages of the irradiation while it intends to allow to determine the conditions (i.e. layer thicknesses, laser beam properties, etc) to fabricate application-based high strength DLMTF. The model proposes a number of issues that a systematic experimental approach is necessary to address in order to validate the theoretical results such as including measuring electron dynamics through reflectivity variation via pump-probe experiments [24]. Furthermore, the revised model can potentially also provide a different perspective in regard to morphological changes on metals as the predominant mechanism of ripple formation is associated with surface plasmon excitation [29-33]; this process is ascribed to the fact that the proposed model can yield different electron temperature values which would vary the resulting ripple periodicity [31]. An extension of the proposed model could be exploited for tailoring the outcome in different classes of materials (dielectrics [34] or semiconductors [35]) where TTM yields an adequate description for irradiation with ultrashort (of moderate pulse duration) pulses. A challenging task would be to incorporate processes that lead to morphological changes in submelting [36] or ablation [31,37] conditions as it is extremely important for emerging applications in the areas of nanomaterials, nanocomposites, nanoelectromechanical devices.

This work was supported by the *'3DNeuroscaffolds'* research project. The author wishes to thank P.A.Loukakos and E.Stratakis for fruitful discussions.

# Supplementary Material

**Thermal response of double-layered metal films after ultrashort pulsed laser irradiation: the role of nonthermal electron dynamics**

George D. Tsibidis[♣]

*Institute of Electronic Structure and Laser (IESL), Foundation for Research and Technology (FORTH), N. Plastira 100, Vassilika Vouton, 70013, Heraklion, Crete, Greece*

1. **Values for the parameters that appear in the dielectric constant expressions**

The values of the parameters that appear in the expressions Eq.2 in the paper which are used to compute the dielectric function of Cu [14] are listed in Tables 1. The electron relaxation time (which is the inverse of the damping frequency $\gamma$ that appears in Eq.2 in the main body of the work) is given by the following expression

$$\tau_e = \frac{1}{B_L T_L + A_e (T_e)^2}, \qquad (SP.1)$$

where $T_e$, $T_L$ are the electron and lattice temperatures, respectively. Values of the coefficients $A_e$, $B_L$ are $1.28\times10^7$ (s$^{-1}$K$^{-2}$) and $1.23\times10^{11}$ (s$^{-1}$K$^{-1}$), respectively, while the procedure which is followed to compute the same values for Ti is described in Section 2. A thorough description of the parameter computation that appear in Table 1 is presented in Ref.1 and references therein.

---

[♣] E-mail: tsibidis@iesl.forth.gr

| | |
|---|---|
| $\varepsilon_\infty$=3.686 | $\Gamma_3$=1.12×10$^{15}$ (rad/sec) |
| $\omega_D$=1.34×10$^{16}$ (rad/sec) | $\Omega_1$=3.205×10$^{15}$ (rad/sec) |
| $B_1$=0.562 | $\Omega_2$=3.43×10$^{15}$ (rad/sec) |
| $B_2$=27.36 | $\Omega_3$=7.33×10$^{15}$ (rad/sec) |
| $B_3$=0.242 | $\Phi_1$=-8.185 |
| $\Gamma_1$=0.404×10$^{15}$ (rad/sec) | $\Phi_2$=0.226 |
| $\Gamma_2$=0.77×10$^{16}$ (rad/sec) | $\Phi_3$=-0.516 |

TABLE 1. Values for the extended Lorentz-Drude model used to compute the dielectric function of Cu [14].

## 2. Computation of the thermo-physical parameters that describes the heat transfer relaxation

To provide an accurate description of the underlying mechanism after irradiation with ultrashort pulses, it is important to treat the thermophysical properties that appear in the model as temperature dependent parameters. To include explicit functions of the thermal parameters on the electron temperature, theoretical data for the two metals (copper and titanium) were calculated.

### (a) Copper

Fig.1S(a,b) illustrate the dependence of the electron heat capacity and electron-phonon coupling constant coefficient on the electron temperature for copper, respectively, while the fitting curves indicate the satisfactory accuracy of the polynomial function. The electron thermal conductivity was calculated by means of a general expression [38]

$$k_e = \frac{\left((\theta_e)^2 + 0.16\right)^{5/4} \left((\theta_e)^2 + 0.44\right)\theta_e}{\left((\theta_e)^2 + 0.092\right)^{1/2} \left((\theta_e)^2 + \eta\theta_L\right)^{5/4}} \chi \quad \text{(SP.2)}$$

$$\theta_e = \frac{T_e}{T_F}, \theta_L = \frac{T_L}{T_F}$$

where, in the case of copper, the parameters that appear in the expression are the Fermi temperature $T_F=8.16\times10^4$K, $\chi=377$Wm$^{-1}$K$^{-1}$=0.139.

**(b) Titanium**

A similar approach is followed for titanium and Fig.2S(a,b) illustrate the accurate fitting for the theoretical data that appear in the work by Lin et al. [39], however, to the best of our knowledge, parameters for the calculation of the electron thermal conductivity is not known. The procedure that is followed to overcome the difficulty is by using alternative expressions [28,40], namely,

$$k_e = k_{e0}\frac{T_e}{\frac{A_e}{B_L}(T_e)^2 + T_L} \qquad (SP.3)$$

$$G = G_0\left(\frac{A_e}{B_L}(T_e + T_L) + 1\right)$$

where the ratio $A_e/B_L$ (see SM Eq.1 ), $k_{e0}$, $G_0$ that are not known can be specified by an appropriate minimisation procedure that ensures that the resulting values for the electron heat capacity and electron-phonon coupling coefficient coincide with the values provided by Lin et al. [39]. The above procedure constitutes a more general approach as in practice the values of the maximum electron temperatures in Titanium (i.e. on the interface) do not exceed 2000K and a polynomial of first order could adequately describe the evolution of the electron heat capacity and electron-phonon coupling coefficient. It has to be noted that SP.2 and SP.3 expressions for thermal electron conductivity are valid for a wide range of electron temperatures (low and high) and in the low limit electron temperature it reduces to the simple

$$k_e = k_{e0}\frac{T_e}{T_L}$$

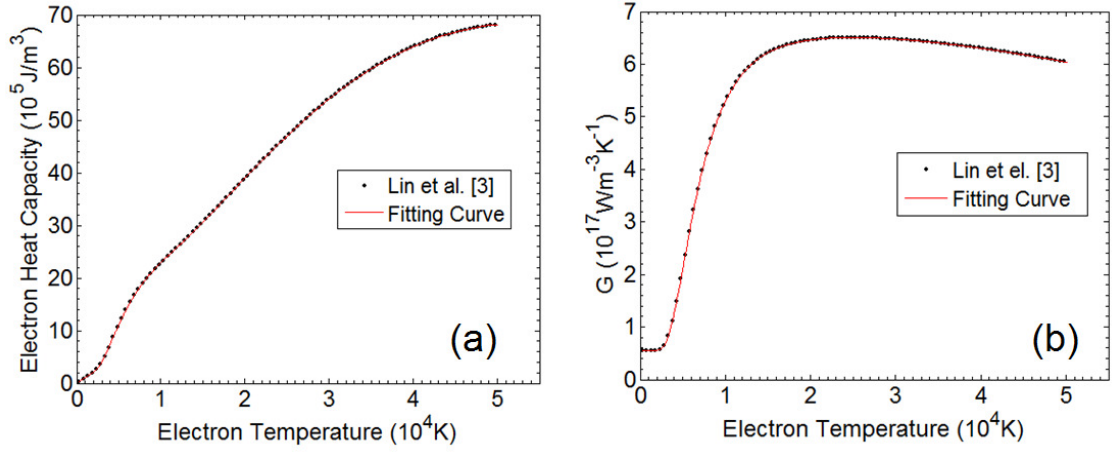

FIG. 1S. Electron heat capacity (a) and electron-phonon coupling coefficient (b) dependence on the electron temperature for copper. Fitting of the theoretical results [39] has been performed by using a polynomial function.

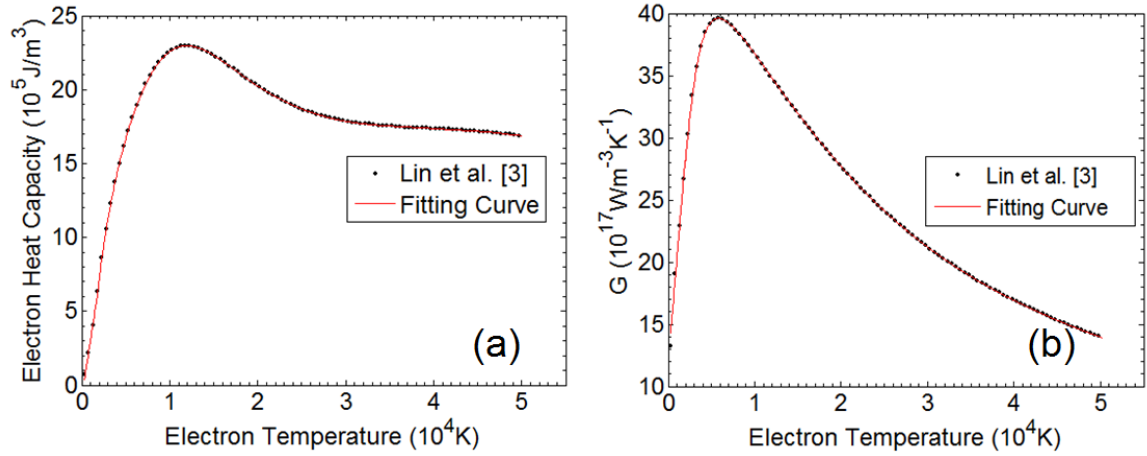

FIG. 2S. Electron heat capacity (a) and electron-phonon coupling coefficient (b) dependence on the electron temperature for titanium. Fitting of the theoretical results [39] has been performed by using a polynomial function.

3. **Computation of the optical properties**

The relationship between the complex refractive index $N$ ($\equiv n-ik$) and the complex dielectric function $\varepsilon$ ($\equiv \varepsilon_1 + i\varepsilon_2$) of a material is given by [41]

$$n(r,z,t) = \sqrt{\frac{\varepsilon_1(r,z,t) + \sqrt{(\varepsilon_1(r,z,t))^2 + (\varepsilon_2(r,z,t))^2}}{2}}$$ (SP.4)

$$k(r,z,t) = \sqrt{\frac{-\varepsilon_1(r,z,t) + \sqrt{(\varepsilon_1(r,z,t))^2 + (\varepsilon_2(r,z,t))^2}}{2}}$$

where $n$, $k$, $\varepsilon_1$, $\varepsilon_2$, are the real (normal) refractive index, extinction coefficient, real and imaginary parts of the dielectric constants of the material, respectively.

To compute the reflectivity on the surface, $R(r,z=0,t)$, and the absorption coefficient $\alpha(r,z,t)$ of the upper layer, the following expressions are used [14,25,42]

$$R(r,z=0,t) = \frac{(1-n)^2 + k^2}{(1+n)^2 + k^2}$$ (SP.5)

$$\alpha(r,z,t) = \frac{2\omega k}{c}$$ (SP.6)

The above expressions are correct to describe reflectivity and absorption when the incident angle (angle between the vertical z-axis and the propagation vector of the beam) and the beam is considered to be *p*-polarised (i.e. polarisation of the electric field is parallel to the plane of incidence and along x-axis). An appropriate modification is required to describe the optical properties in other cases. Furthermore, it was assumed (see description in the main body of the manuscript) that throughout the presented work the thickness of the upper layer is selected to be large enough to assume negligible penetration of the laser beam into the substrate layer and therefore the source term for the second layer approximately vanishes. On the contrary, if the penetration of the beam was considered to be significant, reflection and transmission variations on the interface between the two layers should be taken into account that would result in different expressions for the optical properties [26,43].

4. **Computation of the expressions in the revised model**

The analytic form of $H_{ee}$ and $H_{eL}$ in Eq.3 are given by the following expressions

$$H_{ee}(t-t') = -\frac{\exp\left(-(t-t')\left(h^2v^2/\varepsilon_F^2\tau_0 + 1/\tau_{eL}\right)\right)}{(t-t')^2}$$
$$\times\left[h^2v^2(t-t') + \varepsilon_F^2\tau_0\left(1-\exp\left((t-t')h^2v^2/\varepsilon_F^2\tau_0\right)\right)\right]$$

$$H_{eL}(t-t') = -\frac{\exp\left(-(t-t')\left(h^2v^2/\varepsilon_F^2\tau_0 + 1/\tau_{eL}\right)\right)}{(t-t')\tau_{eL}} \quad \text{(SP.7)}$$
$$\times \varepsilon_F^2\tau_0\left(1-\exp\left((t-t')h^2v^2/\varepsilon_F^2\tau_0\right)\right)$$

where $\varepsilon_F$ is the energy Fermi (7eV for copper [27]), $\tau_0 = 128/\left(\sqrt{3}\pi^2\omega_p\right)$, (with $\omega_p^{(Cu)}$ =1.6454×10$^{16}$ rad/sec, it yields $\tau_0$=0.46fs) [41]. To compute the approximate value for the electron-phonon relaxation time $\tau_{eL}$ for the two metals, the following formula is used

$$\tau_{eL} = \tau_F \frac{hv}{k_B\Theta_D} \quad \text{(SP.8)}$$

where $k_B$=8.621×10$^{-5}$eV/K is the Boltzmann constant, $\Theta_D$ is the Debye temperature (343.5K for Cu [44], $\tau_F$ stands for the time between two subsequent collisions with the lattice calculated using

$$\tau_F = \frac{2.2}{\rho_\mu}\left(\frac{r_s}{a_0}\right)^3 \text{fs} \quad \text{(SP.9)}$$

where $\rho_\mu$ is the resistivity of the material (2.24µΩcm for Cu) [27], $a_0$=0.529×10$^{-8}$ cm is the Bohr radius and $r_s$=2.67$a_0$ for Cu [27] which yields $\tau_F$=18.7fs and $\tau_{eL}$=0.98ps.

5. **Lattice temperature evolution and spatial variation**

To verify the correlation of the influence of the laser irradiation, nonthermal electron interactions with thermal electron and lattice baths, thickness of the upper layer, thermophysical properties of the substrate and heat transfer between the two materials of the double layered film, the spatial variation of the lattice temperature has been sketched. To emphasise on the role of nonthermal electron dynamics, the two models that are compared are: (i) the ETTM, and (ii) the TTM (that assumes a dynamic variation of the optical properties). Fig.3S(a) shows the lattice temperature field at $t$=6ps while Fig.3S(b) illustrates the maximum surface lattice temperature of the system at $r$=2.5μm when the two models are used. It is interesting to discuss the form of the lattice temperature spatial variation and argue on the connection of the variation of the spatial gradient of the temperature with the electron heat conductivity: in principle, the form of the dependence of the lattice temperature should be explained in terms of the interplay between two competing mechanisms:

1. the electron-phonon which induces heat localisation, and

2. the carrier transport (determined by the electron heat conductivity $k_e$) which transfers heat away from the laser-excited region.

Although one might expect that the heat conductivity (Copper's heat conductivity is about twenty times larger than that of Titanium [27]) will play the dominant role in determining the form of the lattice temperature spatial behaviour, it appears that the electron-phonon coupling large discrepancy between the two materials and therefore the increased heat localisation leads to the steep descent of the lattice temperature inside the titanium layer compared with the spatial gradient in the copper region. Similar behaviour has also been observed in other type of bilayered materials (for a similar combination of differences for the electron heat conductivity and electron-phonon coupling) [6,28,45] while the abrupt descent in materials with increasing coupling constant has also been described in single metals [29].

### 6. Transient reflectivity calculation

To validate the proposed theoretical model and compare the theoretical results with experimental observations, we present a short description of the procedure that is required to obtain measurable quantities. Although a more analytical description is part of an ongoing work (as the present paper aims predominantly to present the fundamentals of the thermal response of the doubled layer upon laser irradiation) the basic steps will be presented briefly based on the procedure presented in previous studies [12,16,46]:

To compute electron-thermalisation dynamics, the changes of the electron distribution is related to the modification in the optical properties. Pump-probe experiments are used with a careful choice of the probe wavelength: at probe wavelengths away from the peak of the thermalised response, it is possible to obtain direct observation of the electron thermalisation [12]. Hence, probe wavelengths larger than 830nm (i.e. energies equal to 1.48eV) are expected to provide satisfactory results but other values of the wavelengths that lead to energies around the interband transition (i.e. 2.2 for Cu) could also be investigated to provide a more complete picture of the role of the nonthermal electrons in the thermalisation process. The measured signals (i.e. the transient differential reflection coefficient $\Delta R/R$) can be used to provide a detailed dependence of the optical properties change on the electronic distribution.

A theoretical computation of $\Delta R/R$ can be expressed in terms of the changes of the real and the imaginary parts of the dielectric constant, $\Delta\varepsilon_1$ and $\Delta\varepsilon_2$, respectively, through the following expression

$$\frac{\Delta R}{R} = \frac{\partial \ln R}{\partial \varepsilon_1}\Delta\varepsilon_1 + \frac{\partial \ln R}{\partial \varepsilon_2}\Delta\varepsilon_2 \qquad (SP.10)$$

where SP.4 and SP.5 are employed to compute the derivatives where the probe wavelength is used. As explained in Section 3, if the films are assumed to be very thin and the optical penetration exceeds the thickness of the upper layer, a revision of the reflectivity expression is required.

The change of the imaginary part of the dielectric function is provided by the following expression [46]

$$\Delta\varepsilon_2 = \frac{1}{(\hbar\omega)^2}\int_{E_{min}}^{E_{max}} D(E,\hbar\omega)\Delta f(E,t)dE \qquad (SP.11)$$

where $D(E,\hbar\omega)$ is the joint density of states with respect to the energy $E$ of the final state in the conduction band [12,46]

$$D(E,\hbar\omega) = \frac{1}{(2\pi)^3}\int d^3k\, \delta\!\left[E_c(\vec{k}) - E_d(\vec{k}) - \hbar\omega\right]\delta\!\left[E - E_c(\vec{k})\right] \qquad (SP.12)$$

To compute the differential electron distribution $\Delta f(E,t)$, we have to recall that the laser beam excites electrons from occupied levels below to unoccupied levels above the Fermi energy $\varepsilon_F$ producing a nonthermal step-like change in the electronic distribution that relaxes by both electron-electron and electron-lattice collisions [15]

$$\Delta\rho_{NT}(E,t-t',\vec{r}) = \Delta\rho_i(E,t',\vec{r})\exp\left[-\frac{t-t'}{\tau_0}\left(\frac{E-\varepsilon_F}{\varepsilon_F}\right)^2 - \frac{t-t'}{\tau_{eL}}\right] \quad \text{(SP.13)}$$

where $\Delta\rho_i$ is the initial (i.e. at time $t=t'$) nonthermal electron distribution. The fact that we can probe below or above the Fermi energy indicates that we can attain more information and better insight about electron dynamics. The differential electron distribution is given through the following two expressions [16]

$$\Delta f(E,t) = f(E,t) - 1/\left[1+\exp\left((E-\varepsilon_f)/k_B T_0\right)\right]$$
$$f(E,t) = \Delta A(E,t) + 1/\left[1+\exp\left((E-\varepsilon_f)/k_B T_e\right)\right] \quad \text{(SP.14)}$$

Where $T_0=300K$ and $\Delta A$ is obtained from the integration of SP.13 from $-\infty$ to $t$. Looking at (SP.14), it is obvious that the electron (i.e. the thermalised population) temperature derived from our extended model can provide details about the transient values of the imaginary part of the dielectric constants while $\Delta\varepsilon_1$ is obtained from the Kramers-Kronig relations [27]. The advantage of the proposed model is that it aims to correct the theoretically produced values of transient reflectivity based on a more complete scenario that incorporates contributions from nonthermal electron presence, variation of optical properties and ballistic transport while it aims present a mechanism that is free of fitting constants included in previous approaches [12,13].

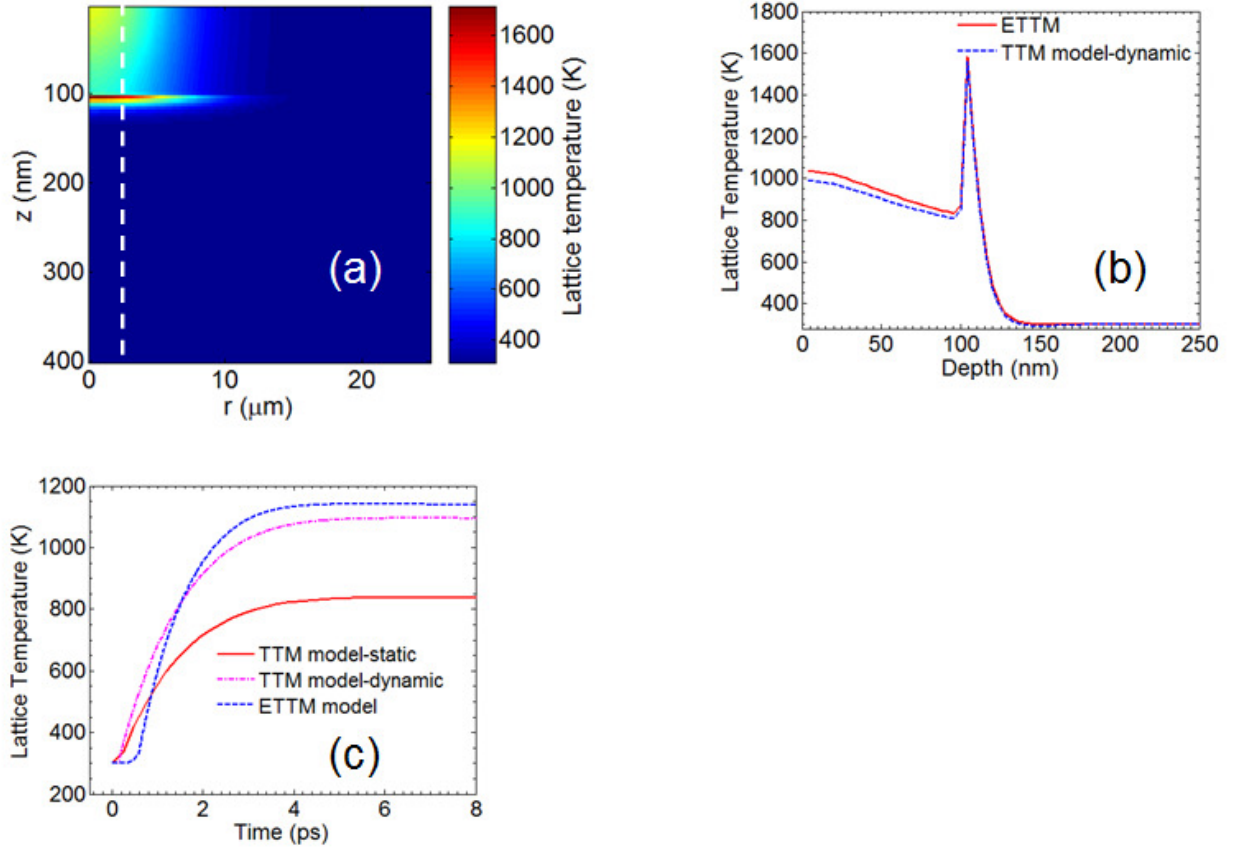

FIG. 3S. (Color Online) (a) Lattice temperature field at $t$=6ps, (b) Comparison of ETTM and TTM (dynamic) for $d_l$=100nm across $r$=2.5μm (*white* line in (a)) at $t$=6ps, , (c) Lattice temperature evolution for the three models at $(r,z)$=(0,0). ($E_p$=0.06J/cm$^2$, $t_p$=50fs, 800nm laser wavelength, $R_0$=10μm).